\documentclass[runningheads]{llncs}

\usepackage[T1]{fontenc}
\usepackage[table]{xcolor}
\usepackage[most]{tcolorbox}
\usepackage{amsmath}
\usepackage{graphicx}
\usepackage{listings}
\usepackage{enumitem}
\usepackage{float}
\usepackage{caption}
\usepackage{subcaption}
\usepackage{booktabs}

\newtcolorbox[auto counter]{ObsBox}{
    borderline west={0pt}{0pt}{white},
    colback=lightgray!30!white,
    boxsep=0pt,           
    left=2pt,             
    right=2pt,            
    breakable
}

\definecolor{codegreen}{rgb}{0,0.6,0}
\definecolor{codegray}{rgb}{0.5,0.5,0.5}
\definecolor{codepurple}{rgb}{0.58,0,0.82}
\definecolor{backcolour}{rgb}{0.95,0.95,0.92}

\lstdefinestyle{mystyle}{
    commentstyle=\color{codegreen},
    keywordstyle=\color{magenta},
    numberstyle=\tiny\color{codegray},
    stringstyle=\color{codepurple},
    basicstyle=\ttfamily\footnotesize,
    breakatwhitespace=false,         
    breaklines=true,                 
    captionpos=b,                    
    keepspaces=true,                 
    numbers=left,                    
    numbersep=5pt,                  
    showspaces=false,                
    showstringspaces=false,
    showtabs=false,                  
    tabsize=2,
    frame=lines
}

\title{Examining MPI and its Extensions for Asynchronous Multithreaded Communication}
\titlerunning{Examining MPI for Asynchronous Multithreaded Communication}

\author{Jiakun Yan\inst{1}\orcidID{0000-0002-6917-5525} \and Marc Snir\inst{1}\orcidID{0000-0002-3504-2468} \and Yanfei Guo\inst{2}\orcidID{0000-0002-3731-5423}}

\authorrunning{J. Yan et al.}

\institute{University of Illinois Urbana-Champaign, Urbana, IL 61801, USA \email{\{jiakuny3,snir\}@illinois.edu} \and
Argonne National Laboratory, Lemont, IL 60439, USA\\
\email{yguo@anl.gov}}

\begin{document}
\maketitle

\begin{abstract}
The increasing complexity of HPC architectures and the growing adoption of irregular scientific algorithms demand efficient support for asynchronous, multithreaded communication. This need is especially pronounced with Asynchronous Many-Task (AMT) systems. This communication pattern was not a consideration during the design of the original MPI specification. The MPI community has recently introduced several extensions to address these evolving requirements. This work evaluates two such extensions, the Virtual Communication Interface (VCI) and the Continuation extensions, in the context of an established AMT runtime HPX. We begin by using an MPI-level microbenchmark, modeled from HPX's low-level communication mechanism, to measure the peak performance potential of these extensions. We then integrate them into HPX to evaluate their effectiveness in real-world scenarios. Our results show that while these extensions can enhance performance compared to standard MPI, areas for improvement remain. The current continuation proposal limits the maximum multithreaded message rate achievable in the multi-VCI setting. Furthermore, the recommended one-VCI-per-thread mode proves ineffective in real-world systems due to the attentiveness problem. These findings underscore the importance of improving intra-VCI threading efficiency to achieve scalable multithreaded communication and fully realize the benefits of recent MPI extensions.
\end{abstract}


\section{Introduction}

High-performance computing (HPC) architectures are becoming increasingly heterogeneous with extensive on-node parallelism. Modern compute nodes often feature over 100 CPU cores and multiple accelerators. Meanwhile, scientific applications are adopting more adaptive or sparse algorithms~\cite{hofmeyr2020metahipmer,GenomePaRSEC2024Ltaief} to achieve higher resolution and scalability. These trends challenge the traditional Bulk-Synchronous Parallel (BSP) model, in which all processes operate in lockstep with evenly distributed workloads. 

Asynchronous Many-Task (AMT) systems have emerged as a compelling alternative. In these systems, applications are expressed as task dependency graphs, and the runtime manages task scheduling, dependencies, and communication. AMT runtimes employ oversubscription, asynchronous execution, and communication-computation overlap to outperform hand-tuned BSP implementations in increasingly irregular workloads~\cite{EarthSystemPaRSEC2024Abdulah,Yadav2023legate_sparse,daiss2024octotiger}.

AMTs exhibit different communication characteristics from BSP applications~\cite{mor2023PaRSEC_LCI,yan2025hpx_lci}. Messages are typically finer-grained and dominated by point-to-point communication rather than global collectives. Communication targets are highly dynamic, with many outstanding operations, and most threads (logically or physically) can generate or consume messages. These characteristics fall outside the traditional design and optimization focus of MPI.

This paper investigates how well existing MPI and recent extensions can support AMT's communication requirements through a case study of an established AMT runtime, HPX. While our focus is on AMTs, their communication challenges are increasingly common in applications with data-dependent execution, beyond the traditional BSP domain. To remain broadly relevant, MPI must evolve to meet these demands.

Building on the analysis of communication requirements of AMT presented in~\cite{yan2025hpx_lci}, we focus on two critical features shown to significantly impact application-level performance: (1) scalable handling of many concurrent communication operations, and (2) effective replication of communication resources to reduce contention. We first use an MPI-level microbenchmark, modeled from HPX's low-level communication mechanism, to evaluate the raw capabilities and limitations of the tested extensions, and then integrate them into HPX to assess their practicality and system-level effectiveness.

Specifically, we evaluate two MPICH extensions:
\begin{itemize}
    \item \emph{MPIX Continuation}: a callback-based completion mechanism designed to reduce overhead from managing large numbers of pending operations.
    \item \emph{MPICH's VCI-mapped communicators}: a mechanism to mitigate thread contention by replicating internal communication resources and mapping them to distinct communicators.
\end{itemize}
Our results reveal both these extensions' advantages and current limitations and motivate recommendations for evolving MPI standards and implementations to better support asynchronous multithreaded runtimes.

The rest of the paper is organized as follows. Section~\ref{sec:background} provides background on the MPI threading model and the extensions we evaluate. Section~\ref{sec:parcelport} describes how we integrate VCI and continuation extensions into the HPX parcelport logic, including the modifications we make to the existing extensions. Section~\ref{sec:pingpong} presents our MPI-level microbenchmark and the fundamental performance characteristics of the extensions. Section~\ref{sec:evaluation} then evaluates the extensions in the context of the HPX runtime, using both microbenchmarks and a real-world astrophysics application, OctoTiger~\cite{daiss2024octotiger}. Finally, Section~\ref{sec:conclusion} concludes the paper and discusses suggestions for improving MPI support for AMT systems.
\section{Background}
\label{sec:background}

\subsection{MPI Threading Level}

The MPI specification~\cite{mpi41} defines four levels of thread support, in increasing order: 
\allowbreak
\texttt{MPI\_THREAD\_SINGLE}, \texttt{MPI\_THREAD\_FUNNELED}, 
\allowbreak
\texttt{MPI\_THREAD\_SERIALIZED}, and 
\texttt{MPI\_THREAD\_MULTIPLE}. \allowbreak
\texttt{MPI\_THREAD\_MULTIPLE} offers the highest level of thread support, allowing multiple threads to invoke MPI functions simultaneously. This model is the most intuitive for writing multithreaded MPI programs and is preferred by many users, according to an MPI survey~\cite{Bernholdt2020MPIsurvey}. Most AMT systems such as HPX~\cite{Kaiser2020HPX}, Legion~\cite{bauer2012LegionExpressingLocality}, and Charm++~\cite{kale1993CHARMPortableConcurrent} rely on \texttt{MPI\_THREAD\_\allowbreak MULTIPLE}. 
However, efficient support for this thread level has historically been lacking in many MPI implementations~\cite{Patinyasakdikul2019multirate}, primarily due to contention on internal MPI data structures and underlying network resources. This work focuses on optimizing and evaluating MPI extensions, specifically in the context of 
\texttt{MPI\_THREAD\_\allowbreak MULTIPLE}.

\subsection{MPICH VCI}

The MPICH Virtual Communication Interface (VCI)~\cite{zambre2020vci} is a mature mechanism in MPICH for addressing the MPI multithreaded efficiency issue through resource replication, emerging from the deprecated MPI endpoint proposal. When enabled, the MPICH runtime will associate a distinct set of communication resources (VCIs) with every MPI communicator, allowing threads to communicate on different communicators with minimal contention. It also has advanced options to map communications to different VCIs according to their communication tags. MPICH recommends multithreaded applications to allocate a separate VCI/communicator for each thread.

The design and implementation of VCI have been covered in detail in~\cite{zambre2020vci,zambre2021LogicallyParallelCommunication}. As a brief overview, a VCI represents a relatively independent set of communication resources needed on the critical path of MPI communication routines. It primarily includes a UCP worker (when using the UCX~\cite{shamis2015ucx} backend) or an OFI domain (when using the OFI~\cite{libfabric} backend), which further encapsulates resources related to network hardware interfacing, memory registration, tag matching, and progressing. MPICH employs a per-VCI spinlock to ensure thread safety, allowing concurrent operations across VCIs but serializing access within each VCI.

\subsection{MPI Continuation Proposal}

The MPI continuation proposal~\cite{mpi_cont} aims to provide an efficient mechanism for managing multiple pending communication operations. In standard MPI, the only way to track the status of pending operations is to wait for or test the request object corresponding to an individual communication operation. However, in event-driven systems such as AMTs, many communication operations may be posted concurrently, and the runtime needs to react when any one of the individual operations completes. 
\texttt{MPI\_Testsome} is unsuitable for this use case as it is typically implemented as a loop over the input request array, and maintaining the request array is inconvenient and expensive. 
Instead, such systems typically maintain lists of MPI requests (i.e., request pools) and use \texttt{MPI\_Test} to opportunistically probe requests in the pool until one or more completed ones are found. A thread polls the pool when it becomes idle~\cite{chatterjee2013IntegratingAsynchronousTask,Mei2011Charm++MPI,yan2023lcipp}.

To avoid the polling overhead and the thread synchronization required to manage shared request pools, the continuation proposal introduces an API that allows MPI clients to attach callback functions to individual requests and register them with a \emph{continuation request}. The application then polls only this continuation request to drive progress, and callbacks are automatically invoked when corresponding communication operations complete. \cite{schuchart2021CallbackbasedCompletionNotification} implements this proposal on a test branch of OpenMPI and integrates it into PaRSEC, where a single thread handles all communication.

In this work, we implement the continuation proposal in MPICH and evaluate it in the context of HPX, where all worker threads can produce and consume messages concurrently. This provides a more realistic test case for multithreaded communication. We further investigate how well the continuation mechanism integrates with multi-VCI configurations and assess its effectiveness in managing completion overhead in these scenarios.
\section{Extend HPX parcelport with MPI Extensions}
\label{sec:parcelport}

In this section, we will describe the design of HPX’s low-level communication layer, known as \emph{parcelport}, and detail how we integrate the two MPI extensions into the parcelport implementation. We also discuss the modifications made to the continuation extension to support multi-VCI scenarios better.

\subsection{Background}

We first briefly describe the HPX communication stack and the original MPI parcelport implementation here. For a complete description of each layer's functionalities and more details, please refer to \cite{yan2025hpx_lci}.

\subsubsection{HPX Communication Stack Overview}

Currently, HPX has three fully functioning communication backends: TCP, MPI, and LCI~\cite{yan2023lcipp}. HPX's communication stack is organized into two layers. The \emph{upper layer} is shared by all backends and handles essential services such as (de)serialization, address resolution, message aggregation, and termination detection. Below it, the \emph{parcelport layer} is backend-specific and implements the actual data transfer mechanism.

In HPX, messages are transmitted in the form of \emph{parcels}, each (logically) consisting of one non-zero-copy (NZC) chunk and an optional set of zero-copy (ZC) chunks. The NZC chunk contains control metadata, while the ZC chunks hold bulk data. Since ZC chunks must be deserialized into memory layouts compatible with C++ data structures, the upper layer needs to pre-allocate appropriate receive buffers before the parcel is fully received (via the \texttt{allocate\_zc\_chunks} function).

Each parcelport must implement two core functions: (1) a non-blocking 
\allowbreak
\texttt{send\_parcel} function to send parcels and invoke a callback when complete, and (2) a \texttt{background\_work} function that checks for incoming parcels and progresses outstanding communication. The \texttt{background\_work} function is frequently invoked by idle threads and notifies the scheduler whether communication made forward progress. It passes the received parcels to the upper layer by calling the \texttt{handle\_parcel} function, which either enqueues the encapsulated tasks or executes them immediately.

\subsubsection{Baseline MPI Implementation}

The original MPI parcelport transfers an HPX parcel using a sequence of MPI messages, consisting of a header followed by one or more data messages. The header contains metadata, such as NZC size, number of ZC chunks, and the MPI tag used for the follow-ups, and may piggyback the NZC chunk if it's small enough. Each remaining chunk is sent in a separate message.

All communication is non-blocking. Header and data messages use \texttt{MPI\_Isend}, with different tags to distinguish them. A single \texttt{MPI\_Irecv} (pre-posted with \texttt{MPI\_ANY\_SOURCE} and the header tag) listens for incoming headers. Upon receiving one, the receiver posts additional \texttt{MPI\_Irecv}`s for the corresponding data messages, using buffer allocations from \texttt{allocate\_zc\_chunks} as needed.

To simplify synchronization, each parcel has at most one active \texttt{MPI\_Isend} or \texttt{MPI\_Irecv} at a time; the following message is posted only after the current one completes. Messages from different parcels may proceed concurrently. The MPI request handles for pending sends and receives (except the preposted receive) are stored in two STL deques (\emph{request pools}). The \texttt{background\_work()} function is responsible for polling the preposted receive request and the request pools using \texttt{MPI\_Test}. The request pools are polled in a round-robin fashion.

Because any HPX worker may enter \texttt{background\_work()}, the parcelport must be thread-safe. MPI is initialized with \texttt{MPI\_THREAD\_MULTIPLE}, and all polling operations are guarded by an HPX lock. Locking uses a non-blocking \emph{try-lock} whenever possible, and waiting user-level threads are descheduled to avoid wasting CPU cycles.

\subsection{Replication of Communicators}

The baseline implementation uses a single communicator, which maps to a single set of internal communication resources and is protected by a single spinlock. This will incur severe thread contention for the lock if multiple threads access it simultaneously for posting sends/receives or testing corresponding requests. The VCI extension in MPICH enables us to replicate internal communication resources by mapping them to distinct communicators. We thus enhance the baseline MPI parcelport with the ability to split communication traffic into a configurable number of communicators. We will call this enhanced parcelport the \emph{MPIx parcelport}.

We must ensure that the send and receive operations for the same MPI message are always posted with the same communicator to guarantee message delivery. Therefore, we construct a static mapping from HPX worker threads to MPI communicators during parcelport initialization. We assign HPX worker threads to communicators in an order that ensures most adjacent threads are assigned the same communicator, thereby improving locality. When the upper layer invokes the \texttt{send\_parcel} function of the parcelport layer on a worker thread, the parcel passed by the \texttt{send\_parcel} function will be associated with the MPI communicator assigned to that thread. The index of the assigned communicator will be passed to the receiver via the header message. All the following MPI send and receive calls for that parcel will be made against the assigned communicator.

With multiple communicators, the MPIx parcelport will prepost one \texttt{MPI\_Irecv} for each communicator for potentially incoming header messages. \emph{background\_work} function will only poll the pre-posted receives associated with the communicator assigned to the worker thread that invokes it. When the continuation extension is not used, the request pools are also replicated per communicator, and the \texttt{background\_work} function will poll the request pools corresponding to the mapped communicator.

The current MPICH implementation employs a hybrid progress model in the case of multiple VCIs: a progress call (happening implicitly inside \texttt{MPI\_Test} and all blocking MPI functions) will primarily progress the VCI that is associated with the calling operation, but it will also progress all VCIs once in a while (every 256 VCI-local progress calls). This provides stronger progress guarantees~\cite{zambre2020HowLearnedStop}, but also increases contention between threads. As a result, we set the \texttt{MPIR\_CVAR\_CH4\_GLOBAL\_PROGRESS} to \texttt{false} to turn off the occasional global progress. Section~\ref{sec:pingpong-gp} analyzes the performance impact of this setting.

\subsection{Replacing Request Polling with Callbacks}

The MPIX Continuation proposal allows clients to attach a callback function to an operation request. In the new MPI parcelport, after we post a \texttt{MPI\_Isend} for a header/follow-up message or a \texttt{MPI\_Irecv} for a follow-up message, we attach a callback function to the resulting request. The callback function will push a completion descriptor to a preallocated completion queue. Essentially, we use the continuation callback to implement a queue-based completion mechanism. The \texttt{background\_work} function will poll the completion queue for any completed operation and react accordingly. We share the completion queue among all threads to improve load balancing. The completion queue uses a highly optimized atomic queue implementation (LCRQ~\cite{Morrison2013lcrq}). The lessons learned from the LCI parcelport show that the atomic completion queue is not a performance bottleneck.

We do not directly invoke the HPX completion logic in the callback because HPX can invoke arbitrary user tasks and even deschedule the current user-level thread, which can lead to reduced performance and even deadlocks. The queue-based design allows us to decouple the upper-level complexity from the low-level communication logic and also improve load balancing.

\subsection{Complication with Continuation Requests}

While the core mechanism of the MPIX Continuation proposal is to attach callback functions to individual MPI operation requests, it is not the entire proposal. To ensure progress and allow more controls over callback execution, the proposal also introduces a persistent \emph{continuation request} object. All continuations (requests with attached callbacks) must be registered with a continuation request. The continuation request is marked complete when all the registered continuations have executed; the continuation request can be tested for completion, and has to be explicitly restarted with \texttt{MPI\_Start} before newly attached continuations can be executed again. In MPICH, an atomic counter per continuation request tracks the total number of pending requests to determine whether the continuation request is complete.

The continuation proposal expects users to test the continuation requests to drive the MPI progress engine. In a multi-VCI setup, the MPI runtime must determine which VCI(s) to make progress on when a continuation request is tested. MPICH adopts the following strategy for selecting the VCI(s) to make progress: each continuation request maintains a per-VCI atomic counter to track the number of pending operations on that VCI; when testing the continuation request, the MPICH implementation will only make progress on the VCI with active associated operations (along with occasional global progress).

In many scenarios, the overhead introduced by the continuation request functionality can be an unnecessary burden: progress can be guaranteed using other MPI calls, and each communication completion already invokes a client-defined callback. From the client's perspective, there is no need to explicitly test for the completion of multiple handler invocations. Therefore, we extend the existing continuation proposal with the option to disable the usage of the continuation request, by setting the \texttt{cont\_request} argument to \texttt{MPI\_REQUEST\_NULL} in the \texttt{MPIX\_Continue} function. In this case, we can avoid the overhead of atomically counting the pending callbacks and completing/restarting the continuation request. We evaluate the performance implications of this optimization in Section~\ref{sec:pingpong-cont_req}.

In HPX, we adopt this optimization and skip the allocation of continuation requests entirely. HPX worker threads periodically poll their pre-posted receives, which automatically invokes the progress engine for the corresponding VCI. This is a lovely coincidence that HPX does not need to do anything additional to ensure the progress of all pending communications attached to continuation callbacks. For other clients where this is not the case, the MPICH runtime provides a non-standard function \texttt{MPIX\_Stream\_progress} to invoke the progress engine of a specific VCI explicitly.

\subsection{Summary}

With the VCI and the continuation extensions, every thread will almost always use its assigned communication resources, including the communicator, preposted receives, and the internal progress engine. The only two exceptions are the shared completion queue and the send/receive of the data messages, which we do not expect to be significant performance bottlenecks.

\section{MPI-level Microbenchmark}
\label{sec:pingpong}

We begin with a multithreaded active message ping-pong microbenchmark to evaluate the basic performance characteristics of the mechanisms used in the MPIx parcelport, independent of the HPX runtime. To do so, we isolate the active message layer from the MPIx parcelport implementation in HPX and construct a standalone microbenchmark on top of it. The benchmark runs on two nodes, each hosting a single MPI process with a configurable number of threads. Threads are pinned to individual cores, and each thread performs a fixed number of ping-pong iterations with a corresponding peer thread on the remote node. All communication is carried out using the active message services provided by the extracted layer. 

The isolated active message layer organizes the relevant MPI resources (including a communicator, a preposted receive request, and a request pool) into a logical unit referred to as a \emph{device}. In the baseline (standard MPI) configuration, all threads share a single device. With the VCI extension enabled, each thread is assigned its device, which is mapped to a VCI. With the continuation extension, the request pool is replaced with callbacks.

\subsection{Experiment Setup}
\label{sec:setup}

We run all the experiments in this section and Section~\ref{sec:evaluation} on SDSC Expanse and NCSA Delta. Table~\ref{table:platform_config} summarizes the platforms' configurations. The two platforms have similar CPUs but have different network hardware and software stacks. Expanse uses HDR InfiniBand with Mellanox ConnectX-6 NICs, while Delta uses HPE Slingshot-11 with HPE Cassini NICs. On InfiniBand, MPICH can use either UCX~\cite{shamis2015ucx} or OFI~\cite{libfabric} as the communication backend, while on Slingshot-11, MPICH can only use OFI. We use a customized version of MPICH 4.3.0 that implements the continuation proposal. This version is currently available in a pull request on the MPICH GitHub repository.

\begin{table}[htbp]
  \caption{Platform Configuration.}
  \label{table:platform_config}
  \begin{center}
  \small
  \begin{tabular}{llll}
  \toprule
  Platform & SDSC Expanse & NCSA Delta \\
  \midrule
   CPU & AMD EPYC 7742 & AMD EPYC 7763 \\
   sockets/node & 2 & 2\\
   cores/socket & 64 & 64 \\
   NIC &  Mellanox ConnectX-6 & HPE Cassini\\
   Network & HDR InfiniBand & Slingshot-11 \\
   & (2x50Gbps) & (200Gbps) \\
   Software & MPICH 4.3.0 & MPICH 4.3.0 \\
   & UCX 1.17.0 & Cray MPICH 8.1.27 \\
   & Libfabric 1.21.0 & Libfabric 1.15.2.0  \\
   & OpenMPI 4.1.3 & SSHOT2.1.3 \\
   & Libibverbs 43.0 & \\
   
  \bottomrule
  \end{tabular}
  \end{center}
\end{table}

\subsection{Overall Performance with Multiple VCIs}

We begin by evaluating the performance impact of using multiple VCIs with different MPICH network backends, and compare the results to those of system-installed MPI implementations (OpenMPI and Cray-MPICH) as well as standard MPICH without VCI extensions.

\begin{figure}[htbp]
    \centering
    \begin{subfigure}{0.6\linewidth}
        \includegraphics[width=\linewidth]{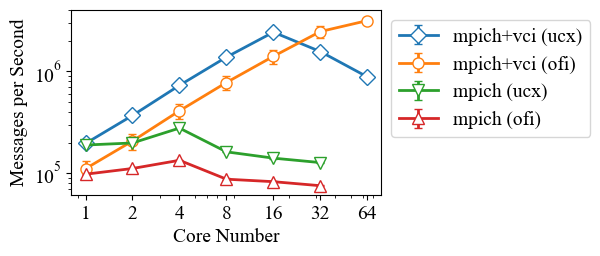}
        \caption{Experiment Results on Expanse with 1-64 threads per process.}
    \end{subfigure}
    \begin{subfigure}{0.6\linewidth}
        \includegraphics[width=\linewidth]{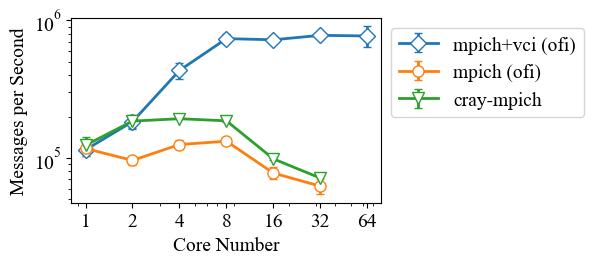}
        \caption{Experiment Results on Delta.}
    \end{subfigure}
    \caption{Performance impacts of the VCI extension compared to other MPI variants with 1-64 threads per process.}
    \label{fig:pingpong-vci}
\end{figure}

As shown in Fig.~\ref{fig:pingpong-vci}, the MPICH VCI extension significantly improves the multithreaded performance of MPI, outperforming both the system-installed MPI (OpenMPI and Cray-MPICH) and standard MPICH itself by many-fold. When comparing the best-performing multi-VCI configurations against the best standard MPI configurations using 64 threads per process, we observe speedups of 15x on Expanse and 8x on Delta. However, the performance gain depends on the underlying network backend, revealing a trade-off between UCX and OFI. While UCX has better base performance, it scales poorly when the number of threads/VCIs exceeds 16. On Expanse with 64 threads (and 64 VCIs), MPICH with the OFI backend outperforms its UCX counterpart by a factor of 4×.

In the standard MPI configuration, all threads share a single device (i.e. a communicator, a preposted receive, and a request pool). For comparison, we also evaluated a variant where each thread has its own device, still using standard MPI. However, it results in even lower performance than the shared device case. This is because with multiple outstanding pre-posted receive requests, the MPI has more chances to contend for the blocking locks inside the VCI. 

We have also compared the performance of the continuation extension against plain request polling. However, we found no performance difference between the two approaches in this microbenchmark. This is expected, as in this ping-pong microbenchmark, each thread has only one send request and one receive request to poll simultaneously.

\subsection{Global Progress with Multiple VCIs}
\label{sec:pingpong-gp}

\begin{figure}
    \centering
    \begin{subfigure}{0.49\linewidth}
        \includegraphics[width=\linewidth]{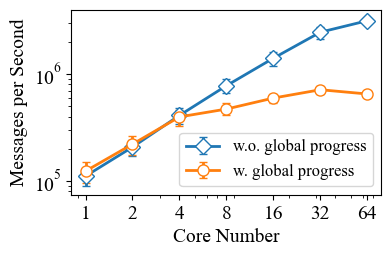}
        \caption{Expanse.}
    \end{subfigure}
    \begin{subfigure}{0.49\linewidth}
        \includegraphics[width=\linewidth]{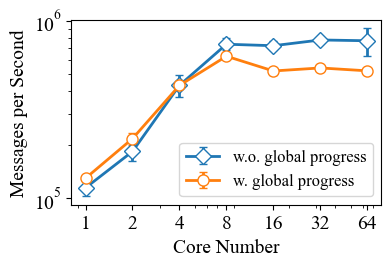}
        \caption{Delta.}
    \end{subfigure}
    \caption{Performance impacts of the global progress requirement with 1-64 threads per process.}
    \label{fig:pingpong-gp}
\end{figure}

Figure~\ref{fig:pingpong-gp} shows the performance impact of the occasional global progress inside MPICH. We evaluate two variants (configured by the \texttt{MPIR\_CVAR\_CH4\_GLOBAL\_PROGRESS} control variable): one with occasional global progress enabled (the default option) and the other with it disabled (the option used by HPX). We observe that performance is significantly improved when we disable the global progress option, even though it only performs one global progress every 255 per-VCI progress updates. The message rate is improved by 5x in the case of Expanse and 40\% in the case of Delta.

\subsection{Continuation with Multiple Threads}
\label{sec:pingpong-cont_req}

\begin{figure}
    \centering
    \begin{subfigure}{0.49\linewidth}
        \includegraphics[width=\linewidth]{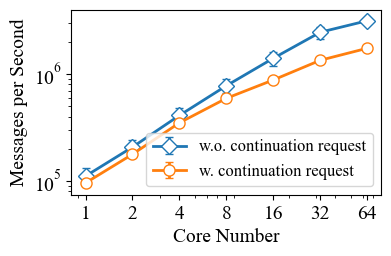}
        \caption{Expanse.}
    \end{subfigure}
    \begin{subfigure}{0.49\linewidth}
        \includegraphics[width=\linewidth]{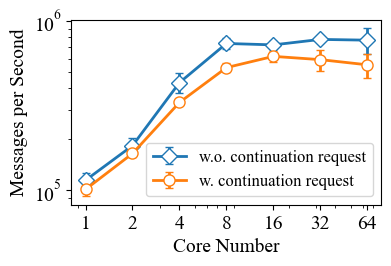}
        \caption{Delta.}
    \end{subfigure}
    \caption{Performance impacts of the continuation request with 1-64 threads per process.}
    \label{fig:pingpong-cont}
\end{figure}

Figure~\ref{fig:pingpong-cont} shows the performance impact of the continuation request. We evaluate two variants with/without the continuation requests. The variant with the continuation request allocates one continuation request per VCI, so there will be no contention on the VCI progress engines. The performance is improved when we disable the continuation request (by passing \texttt{MPI\_REQUEST\_NULL} as the \emph{cont\_request} argument to the \texttt{MPIX\_Continue} function). The performance is improved by 78\% in the case of Expanse and 27\% in the case of Delta.

\subsection{Summary}
The VCI extension greatly improves the maximum message rate achievable in multithreaded scenarios. However, the two existing network backends in MPICH (UCX and OFI) both have limitations. The global progress requirement of the MPI specification and the continuation request construct of the existing continuation proposal also hurt the performance.
\section{HPX Evaluation}
\label{sec:evaluation}

In this section, we evaluate the performance impacts of the VCI and continuation extensions on the new MPI parcelport. 

We use two major benchmarks to evaluate the performance of the new MPI parcelport and the effects of relevant MPICH extension: an HPX microbenchmark with a flood of messages between two nodes testing the maximum throughput of message processing; and an astrophysics application, OctoTiger~\cite{daiss2024octotiger}, testing the impact on a real-world application. The details of the HPX flooding microbenchmark can be found in \cite{yan2025hpx_lci}.

For the HPX microbenchmark, we show the message rate achieved with two message sizes: 8 bytes and 16 kilobytes. With 8-byte messages, the header message can piggyback the application data, and every parcel is translated into one low-level MPI message. With 16-kilobyte messages, the data message is too large to piggyback, so every parcel is translated into two MPI messages: one header message and one data message. For the OctoTiger benchmark, we show the total execution time of the application with 20 iterations on 32 nodes. Every HPX process has 63 HPX threads, reserving 1 CPU core for OS activities.

\subsection{Overall Performance}

We first compare the new MPI parcelport (\emph{mpix}) with the existing LCI parcelport (\emph{lci}) and the original MPI parcelport (\emph{mpi}) in HPX.

\begin{figure}[htbp]
    \centering
    \begin{subfigure}{0.49\linewidth}
        \includegraphics[width=\linewidth]{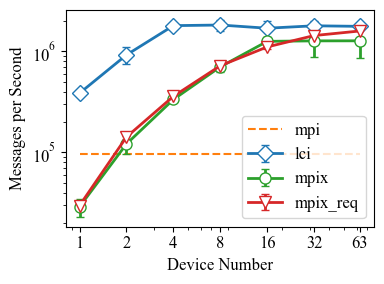}
        \caption{Message Rate (8B) achieved with the flooding microbenchmark on Expanse.}
    \end{subfigure}
    \begin{subfigure}{0.49\linewidth}
        \includegraphics[width=\linewidth]{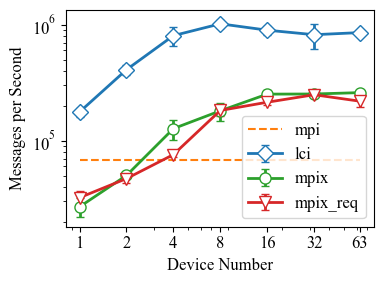}
        \caption{Message Rate (8B) achieved with the flooding microbenchmark on Delta.}
    \end{subfigure}
    \begin{subfigure}{0.49\linewidth}
        \includegraphics[width=\linewidth]{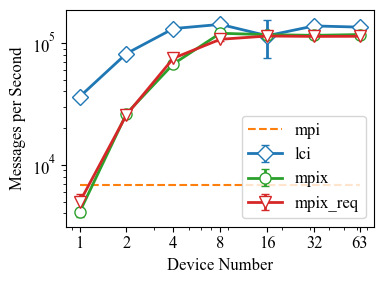}
        \caption{Message Rate (16KiB) achieved with the flooding microbenchmark on Expanse.}
    \end{subfigure}
    \begin{subfigure}{0.49\linewidth}
        \includegraphics[width=\linewidth]{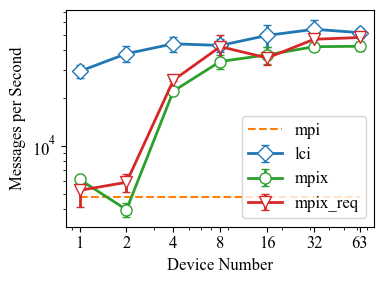}
        \caption{Message Rate (16KiB) achieved with the flooding microbenchmark on Delta.}
    \end{subfigure}
    \begin{subfigure}{0.49\linewidth}
        \includegraphics[width=\linewidth]{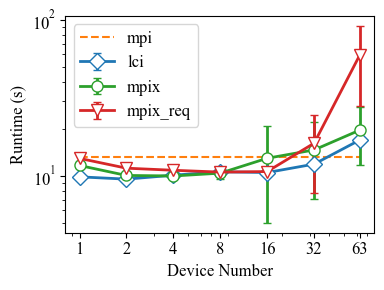}
        \caption{Octo-Tiger time per step with 32 nodes on Expanse.}
    \end{subfigure}
    \begin{subfigure}{0.49\linewidth}
        \includegraphics[width=\linewidth]{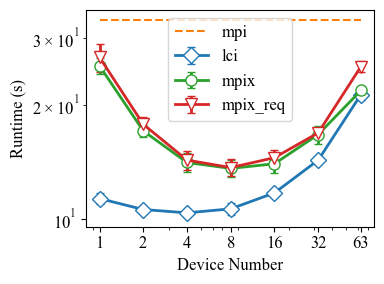}
        \caption{Octo-Tiger time per step with 32 nodes on Delta.}
    \end{subfigure}
    \caption{Performance impacts of using multiple VCIs and continuation. \emph{lci} uses LCI. \emph{mpi} uses the original MPI parcelport. \emph{mpix} uses MPICH with the VCI and continuation extensions.}
    \label{fig:config-ndevices}
\end{figure}

Fig.~\ref{fig:config-ndevices} shows the experimental results of the three benchmarks with different numbers of MPICH VCIs. We show the results of the LCI parcelport (\emph{lci}), the new MPI parcelport with continuation (\emph{mpix}), the new MPI parcelport with request polling (\emph{mpix\_req}), and the old MPI parcelport (\emph{mpi}). We observe that \emph{mpix} greatly shrinks the performance gap between \emph{lci} and \emph{mpi}, especially on Expanse when the device number is large. The performance of \emph{mpi} is much worse than that of \emph{mpix}, showing the performance benefit of the VCI extensions.

Continuation-based programming is simpler than storing multiple requests in a request pool and polling them periodically, so the continuation extension has value in terms of programmability. It also shows a 5\% performance improvement for OctoTiger on Expanse, but failed to demonstrate its performance benefits on other scenarios. This is against what we expected. It shows that the request polling overhead associated with request pools is not as significant as we thought.

We should note that prior study of the LCI parcelport \cite{yan2025hpx_lci} observed that a lightweight polling mechanism is indeed beneficial compared to the request polling mechanism, seemingly contradicting the observation here. However, LCI has a more thread-efficient runtime than MPICH, as LCI uses atomic-based data structures while MPICH uses a per-VCI spinlock to ensure thread safety. As a result, the lock-based request polling mechanism indeed leads to more severe thread contention in the LCI parcelport, but this effect is hidden in MPICH as MPICH's per-VCI spinlock is already coarse-grained. As a result, we believe the continuation extension will be beneficial in the future when the MPICH runtime gets rid of the coarse-grained per-VCI spinlock and uses a more efficient lock-free data structure.

\subsection{Investigate the Slowdown with Too Many VCIs}

The conventional wisdom is that using one VCI per thread will lead to the best performance for multithreaded applications. However, we find that using too many VCIs can lead to performance degradation in real-world applications. We have observed this in the Octo-Tiger benchmark, where performance deteriorates when more than 16 VCIs are used. We could see a similar upward curve with the LCI parcelport. We further investigated why too many MPICH VCIs/LCI devices worsen performance. We identify \textbf{the attentiveness problem} as the main reason.

With too many VCIs, each VCI may not get enough attention from the threads. For 63 threads and 63 VCIs, each VCI only gets one thread to poll it. If the thread gets stuck executing a long-running task, its corresponding VCI will not be polled, and the pending requests on that VCI will not be processed, even though many threads are idle and waiting for work. On the contrary, if there are fewer VCIs, each VCI will get more threads to poll it, and the pending requests on that VCI will be processed more quickly. On the other hand, there is more contention with fewer VCIs.

To demonstrate this assumption and explore a potential fix, we implement a new progress strategy (\emph{random}) in the new MPI parcelport and the LCI parcelport.  In the \emph{random} strategy, each thread randomly picks a VCI to poll from all available VCIs. This way, even if a thread gets stuck in executing a long-running task, other threads can still progress the pending requests on that VCI. Correspondingly, we name the previous strategy as \emph{local}, as each thread only polls its own local VCI.

\begin{figure}[htbp]
    \centering
    \begin{subfigure}{0.49\linewidth}
        \includegraphics[width=\linewidth]{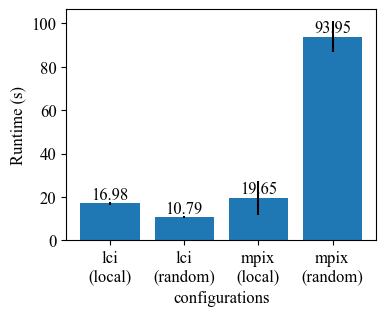}
        \caption{Expanse.}
    \end{subfigure}
    \begin{subfigure}{0.49\linewidth}
        \includegraphics[width=\linewidth]{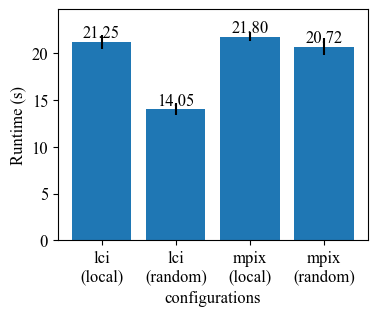}
        \caption{Delta.}
    \end{subfigure}
    \caption{Performance impacts of the \emph{random} progress strategies on OctoTiger execution time with 63 threads/VCIs per process.}
    \label{fig:config-attentiveness}
\end{figure}

Figure~\ref{fig:config-attentiveness} shows the performance impact of the \emph{random} progress strategy. We notice that it greatly improves the performance of the LCI parcelport. However, it does not improve the performance of the new MPI parcelport. Instead, it makes the performance worse. This is due to the different threading efficiency of the two communication runtimes. In MPICH, every progress call will block waiting for the per-VCI spinlock, while in LCI, the progress call is non-blocking and always employs a try-lock wrapper around the low-level network resources~\cite{yan2025lci}. Profiling confirms that MPICH gets stuck in the VCI spinlock more often with the \emph{random} strategy.

\subsection{Summary}
The VCI extension shows great performance benefits across HPX microbenchmarks and a real-world application. The continuation extension does not show much performance benefits compared to per-VCI request polling with the current MPICH implementation. The recommended usage of one VCI per thread does not work with real-world applications that have both computation and communication tasks due to the attentiveness problem.
\section{Related Work}
\label{sec:related_work}

Multiple efforts have sought to improve MPI performance in multithreaded environments. Prior work~\cite{balaji2008EfficientSupportMultithreaded,dozsa2010EnablingConcurrentMultithreaded,amer2015MPIThreadsRuntime,amer2019LockContentionManagement,patinyasakdikul2019GiveMPIThreading} has focused on reducing lock contention and minimizing the scope of critical sections within the MPI runtime. Other approaches \cite{huang2003adaptiveMPI,kamal2010FG-MPI,carribault2010mpc-mpi,hori2018pip-mpi} leverage user-level threads, task systems, or process-in-process techniques to enhance MPI on many-core processors and irregular workloads. More recently, 
 \cite{zambre2020HowLearnedStop,zambre2021LogicallyParallelCommunication} proposed using VCI to replicate low-level network resources, thereby removing the need for runtime-level serialization. The VCI mechanism has since become the recommended approach for improving multithreaded performance in MPICH, representing a major milestone in MPI implementation-level optimization.

In parallel with implementation improvements, a complementary line of research has focused on extending the MPI interface to better support multithreaded execution. ~\cite{dinan2014endpoints} proposed the endpoints extension, which decouples threads from ranks and enables threads within a process to independently issue MPI calls with different endpoints. 
More recent MPIX Stream \cite{zhou2022mpix_stream} and the thread communicator extension \cite{zhou2023thread_comm} revive and refine the endpoint model. These interface-level extensions are designed to help users better convey thread-level parallelism to the MPI runtime. Under the hood, the MPI runtime still relies on VCI-like optimization to provide better multithreaded performance. 

Beyond the MPI ecosystem, several other communication libraries have been developed to support asynchronous and multithreaded communication. GASNet and GASNet-EX~\cite{bonachea2002gasnet,bonachea_gasnet-ex_2018} provide low-level active messages and RMA operations for library developers and compiler-generated codes. At a higher level, PGAS models like UPC~\cite{el2006upc}, UPC++~\cite{bachan2019upc++}, and OpenSHMEM~\cite{chapman2010openshmem} expose global memory abstractions with one-sided RMA operations. LCI~\cite{yan2025lci} proposes new interface and runtime designs to further enhance multithreaded communication performance and simplify asynchronous programming.

In contrast to these efforts, our work does not propose new MPI extensions or communication libraries. Instead, we focus on evaluating the practical effectiveness of existing mechanisms—specifically, the VCI and continuation extensions—and identifying their limitations. Our analysis is complementary to prior work, offering detailed insights into how current MPI features can be better utilized and where future improvements are needed.
\section{Conclusion and Discussion}
\label{sec:conclusion}

In this paper, we evaluated the effectiveness of the VCI and continuation extensions in MPICH using both microbenchmarks and HPX. Our results show that the VCI extension can significantly improve the performance of multithreaded applications. The continuation extension, while beneficial for programmability, currently shows limited performance benefit.

Contrary to the common recommendation of assigning one VCI per thread, we found that excessive use of VCIs can degrade performance in real-world applications. We identified the attentiveness problem as the primary cause: when too many VCIs are in use, the MPI runtime may fail to poll them frequently enough, leading to increased latency and missed progress opportunities. Our findings highlight intra-VCI threading efficiency as a critical factor. Improving it not only resolves the attentiveness issue by enabling more efficient polling across threads, but also allows users to meet their multithreaded communication needs with fewer VCIs—boosting scalability by reducing resource overhead.

Improved intra-VCI efficiency also helps demonstrate the benefits of the continuation extension. Continuations eliminate the need for explicit polling of shared request pools, thus removing the associated thread contention. However, if intra-VCI operations rely on coarse-grained locks, internal contention can obscure these gains. With more efficient intra-VCI handling, continuations can better realize their potential of minimizing overhead and avoiding contention.

While it is known to be challenging to design a threading-efficient VCI due to the non-overtaking requirement and the need to support wildcard receives, recent MPI info keys such as \emph{allow\_overtaking} and \emph{no\_any\_tag/source} offer a practical path forward. When these keys are set, MPI runtimes can safely adopt more scalable designs. Task systems — some of the primary users of asynchronous multithreaded communication — can often tolerate these relaxations~\cite{yan2025hpx_lci}. However, they still require support for \emph{any\_source} receives, which may necessitate additional info keys. One possible approach is to propagate \emph{any\_source} information to the sender side, as suggested in prior work~\cite{Vu2016millionthreads}.

In addition, our evaluation has revealed limitations in two commonly used communication middlewares: UCX and libfabric. Specifically, UCX shows performance degradation when more than 16 UCP workers are used, and libfabric delivers lower absolute performance. Prior work on LCI~\cite{yan2025lci} has demonstrated that multithreaded performance comparable to MPI-everywhere (one process per core) is achievable when building directly on top of the libibverbs layer. Addressing these performance constraints in the underlying middleware is essential for MPI implementations to fully realize scalable multithreaded communication.

We believe these insights offer practical guidance for improving multithreaded communication performance in MPICH and other MPI implementations, and we hope they inform future runtime and interface design.
\subsubsection{Acknowledgements.}
This work used Expanse at San Diego Supercomputer Center~\cite{Strande2021expanse} and Delta at National Center for Supercomputing Applications~\cite{gropp2023delta} through allocations CCR130058 and CIS250465 from the Advanced Cyberinfrastructure Coordination Ecosystem: Services \& Support (ACCESS) program~\cite{boerner2023access}, which is supported by U.S. National Science Foundation grants \#2138259, \#2138286, \#2138307, \#2137603, and \#2138296.

\end{document}